\documentclass[12pt]{article}
\usepackage{amssymb}
\makeatletter
\def\numberbysection{\@addtoreset{equation}{section}
 	\def\theequation{\thesection.\arabic{equation}}}
\makeatother

\numberbysection

\newcommand{\be}{\begin{eqnarray}}
\newcommand{\ee}{\end{eqnarray}}
\newcommand{\non}{\nonumber}
\newcommand{\tr}{\mathop{\rm tr}\nolimits}
\newcommand{\sn}{\mathop{\rm sn}\nolimits}
\newcommand{\cn}{\mathop{\rm cn}\nolimits}
\newcommand{\dn}{\mathop{\rm dn}\nolimits}
\newcommand{\kk}{\kappa}
\newcommand{\id}{\mathbb{I}}

\begin{document}

\begin{titlepage}
\strut\hfill UMTG--231
\vspace{.5in}
\begin{center}

\LARGE Bethe Ansatz solution of the open XX spin chain with 
nondiagonal boundary terms\\[1.0in]
\large Rafael I. Nepomechie\\[0.8in]
\large Physics Department, P.O. Box 248046, University of Miami\\[0.2in]  
\large Coral Gables, FL 33124 USA\\

\end{center}

\vspace{.5in}

\begin{abstract}
We consider the integrable open XX quantum spin chain with nondiagonal
boundary terms.  We derive an exact inversion identity, using which we
obtain the eigenvalues of the transfer matrix and the Bethe Ansatz
equations.  For generic values of the boundary parameters, the Bethe
Ansatz solution is formulated in terms of Jacobian elliptic functions.
\end{abstract}

\end{titlepage}

\setcounter{footnote}{0}

\section{Introduction}\label{sec:intro}

We consider in this article the open XX quantum spin chain with 
nondiagonal boundary terms, defined by the Hamiltonian
\be
{\cal H }&=& {1\over 2}\Big\{ \sum_{n=1}^{N-1}\left( 
\sigma_{n}^{x}\sigma_{n+1}^{x}+\sigma_{n}^{y}\sigma_{n+1}^{y}\right)
\non \\
&+& i \coth \xi_{-} \sigma_{1}^{z}
+ {2 i \kk_{-}\over \sinh \xi_{-}}\sigma_{1}^{x} 
-i \coth \xi_{+} \sigma_{N}^{z}
+ {2 i \kk_{+}\over \sinh \xi_{+}}\sigma_{N}^{x} \Big\} \,,
\label{Hamiltonian}
\ee
where $\sigma^{x} \,, \sigma^{y} \,, \sigma^{z}$ are the standard
Pauli matrices, and $\xi_{\pm} \,, \kk_{\pm}$ are arbitrary boundary
parameters.  This model is known to be integrable \cite{Sk, dVGR, GZ}. 
It has been investigated \cite{Gu, Bi} using a fermionization
technique \cite{JW, LSM}, suitably adapted to accommodate boundary
terms \cite{Tj, BP}.  However, this model has until now resisted a
direct Bethe Ansatz solution due to the absence of a simple reference
(pseudovacuum) state.  \footnote{For the special case of diagonal
boundary terms (i.e., $\kk_{\pm}=0$), a simple pseudovacuum state does
exist, and a Bethe Ansatz solution is known \cite{ABBBQ, Sk}.} Such a
solution is desirable for a number of reasons.  First, the open XX
chain is a special case of integrable open XXZ and open XYZ chains,
which should also admit Bethe Ansatz solutions but which cannot be
solved by fermionization.  Second, Bethe Ansatz solutions are
particularly well-suited for investigating physical properties, such
as ground state, low-lying excitations, scattering matrices, etc.  In
particular, the Bethe Ansatz approach avoids the projection mechanism
\cite{Bi} which can be implemented only in special cases.  (See
\cite{Gu, Bi} and references therein for discussions of interesting
physical applications of the open XX spin chain.)  Finally, the
Sklyanin transfer matrix for the open XX spin chain is closely related
to the Yang matrix \cite{Ya, FS, AN} for a large class of integrable
${\cal N}=2$ supersymmetric quantum field theories with boundary
\cite{FI, Ne}.  Diagonalization of this matrix is a key step in
formulating the thermodynamic Bethe Ansatz equations for these ${\cal
N}=2$ supersymmetric models.

In this paper we derive an exact inversion identity for the model
(\ref{Hamiltonian}), using which we obtain the eigenvalues of the
transfer matrix and the Bethe Ansatz equations.  (This approach does
not completely circumvent the problem of not having a reference state,
since the eigenvectors are not determined.)  We obtain the inversion
identity using the open-chain fusion formula \cite{MN1} together with
the remarkable fact that, for the open XX spin chain/ 6-vertex
free-Fermion model, the fused transfer matrix is proportional to the
identity matrix.  A similar strategy has recently been used \cite{AN}
to solve the open 8-vertex free-Fermion \cite{FW} model, which
corresponds to the case of ${\cal N}=1$ supersymmetry.  These
techniques are generalizations of those which have been developed for
closed spin chains \cite{Ba, Fe, fusion, KS}.

Even though the transfer matrix is constructed entirely from 
hyperbolic functions, we find that the Bethe Ansatz solution is 
formulated in terms of Jacobian elliptic functions for generic values 
of the boundary parameters.  (In contrast, for the XYZ chain 
\cite{Ba}, such functions appear already in the transfer matrix.) For 
special values of the boundary parameters, the elliptic functions 
degenerate into ordinary hyperbolic or trigonometric functions.

The outline of this paper is as follows.  In Section 2, we review the
construction of the Sklyanin transfer matrix for the open XX chain, 
and derive some of its important properties. 
We derive in Section 3 the inversion identity, which we then use in
Section 4 to determine the Bethe Ansatz solution.  In Section 5, we
investigate some special cases in which the solution can be expressed
in terms of ordinary hyperbolic functions.  In particular, we verify
that our solution is similar to the known one \cite{ABBBQ, Sk} for the
case of diagonal boundary terms.  In Section 6, we conclude with a
brief discussion of some possible directions for future work.

\section{Transfer matrix}\label{sec:transfer}

The object of central importance in the construction of integrable 
quantum spin chains is the one-parameter family of commuting matrices 
called the transfer matrix. The transfer matrix for an open chain is 
made from two basic building blocks, called $R$ (bulk) and $K$ 
(boundary) matrices.

An $R$ matrix is a solution of the Yang-Baxter equation 
\be
R_{12}(u-v)\ R_{13}(u)\ R_{23}(v)
= R_{23}(v)\ R_{13}(u)\ R_{12}(u-v) \,.
\label{YB}
\ee 
(See, e.g., \cite{KS, KBI, NePrimer}.) The XX spin chain is a special 
case of the XXZ spin chain, corresponding (in the notation of 
\cite{Sk}) to the anisotropy value $\eta = {i \pi\over 2}$.
The $R$ matrix is therefore the $4 \times 4$ matrix
\be
R(u) = \left( \begin{array}{cccc}
	a             &0         &0           &0            \\
        0             &b         &c           &0            \\
	0             &c         &b           &0            \\
	0             &0         &0           &a 
\end{array} \right) \,, 
\label{bulkRmatrix}
\ee 
with matrix elements
\be
a  = \sinh  (u + {i \pi\over 2})
\,, \qquad
b   = \sinh  u
\,, \qquad
c= \sinh {i \pi\over 2}  \,,
\label{Rmatrixelements}
\ee
which satisfy the free-Fermion condition
\be
a^{2} + b^{2} = c^{2} \,.
\ee
This $R$ matrix has the symmetry properties
\be
R_{12}(u) = {\cal P}_{12} R_{12}(u) {\cal P}_{12} 
= R_{12}(u)^{t_{1} t_{2}} \,,
\ee
where ${\cal P}_{12}$ is the permutation matrix and $t$ denotes 
transpose.  Moreover, it satisfies the unitarity relation
\be
R_{12}(u)\ R_{12}(-u) = \zeta(u) \id \,, \qquad 
\zeta(u)=-\cosh^{2} u \,,
\label{unitarity}
\ee
and the crossing relation 
\be
R_{12}(u) = V_{1} R_{12}(-u-\rho)^{t_{2}} V_{1} \,,
\label{crossing}
\ee 
with 
\be
\rho = - {i \pi\over 2} \,, \qquad V =  \sigma^{x} \,.
\ee 
Finally, it has the periodicity property
\be
R_{12}(u+i \pi)= - \sigma^{z}_{2} R_{12}(u) \sigma^{z}_{2} =
- \sigma^{z}_{1} R_{12}(u) \sigma^{z}_{1} \,.
\label{periodicityR}
\ee 

The matrix $K^{-}(u)$ is a solution of the boundary Yang-Baxter equation 
\cite{Ch}
\be
R_{12}(u-v)\ K^{-}_{1}(u)\ R_{21}(u+v)\ K^{-}_{2}(v)
= K^{-}_{2}(v)\ R_{12}(u+v)\ K^{-}_{1}(u)\ R_{21}(u-v) \,.
\label{boundaryYB}
\ee 
We consider here the following $2 \times 2$ matrix \cite{dVGR, GZ}
\be
K^{-}(u) = \left( \begin{array}{cc}
\sinh(\xi_{-} + u)   & \kk_{-} \sinh  2u \\
\kk_{-} \sinh  2u     & \sinh(\xi_{-} - u) 
\end{array} \right) \,,
\label{Kminusmatrix}
\ee 
which evidently depends on two boundary parameters $\xi_{-} \,, 
\kk_{-}$.  We set the matrix $K^{+}(u)$ to be $K^{-}(-u-\rho)$ with 
$(\xi_{-} \,, \kk_{-})$ replaced by $(\xi_{+} \,, \kk_{+})$; i.e.,
\be
K^{+}(u) = \left( \begin{array}{cc}
i\cosh(\xi_{+} - u)   &  \kk_{+} \sinh  2u \\
\kk_{+} \sinh  2u    & -i\cosh(\xi_{+} + u) 
\end{array} \right) \,,
\label{Kplusmatrix}
\ee 
We shall often use an alternative \cite{GZ} set of boundary parameters
$(\eta_{\mp} \,, \vartheta_{\mp})$ which is related to the set
$(\xi_{\mp} \,, \kk_{\mp})$ by
\be
\cos \eta_{\mp} \cosh \vartheta_{\mp} = 
{i\over 2\kk_{\mp}}\sinh \xi_{\mp} \,, \qquad
\cos^{2} \eta_{\mp} + \cosh^{2} \vartheta_{\mp} = 
1 + {1\over 4\kk^{2}_{\mp}} \,.
\label{alternative}
\ee
The $K$ matrices have the periodicity property
\be
K^{\mp}(u + i \pi) = - \sigma^{z} K^{\mp}(u)  \sigma^{z} \,.
\label{periodicityK}
\ee 

The transfer matrix $t(u)$ for an open chain of $N$ spins is given by 
\cite{Sk}
\be
t(u) = \tr_{0} K^{+}_{0}(u)\  
T_{0}(u)\  K^{-}_{0}(u)\ \hat T_{0}(u)\,,
\label{transfer}
\ee
where $\tr_{0}$ denotes trace over the ``auxiliary space'' 0,
and $T_{0}(u)$, $\hat T_{0}(\lambda)$ are so-called monodromy 
matrices \footnote{As is customary, we usually suppress the 
``quantum-space'' subscripts $1 \,, \ldots \,, N$.}
\be
T_{0}(u) = R_{0N}(u) \cdots  R_{01}(u) \,,  \qquad 
\hat T_{0}(u) = R_{10}(u) \cdots  R_{N0}(u) \,. 
\label{monodromy}
\ee
Indeed, Sklyanin has shown that $t(u)$ constitutes a one-parameter 
commutative family of matrices
\be
\left[ t(u)\,, t(v) \right] = 0  \,.
\label{commutativity}
\ee 
Typically, the Hamiltonian is proportional to the first derivative
of the transfer matrix $t'(0)$ \cite{Sk}.  However, this quantity is
trivial for the XX model, due to the fact $\tr K^{+}(0)=0$.  In order
to obtain the Hamiltonian, we must go to the second derivative
\cite{CGR}.  We find
\be
{\cal H} &=& {t''(0)\over 4 (-1)^{N+1} i \sinh \xi_{-} \tr 
K^{+'}(0)}\non  \\
&=& \sum_{n=1}^{N-1}H_{n\,, n+1} + {i\over 2 \sinh \xi_{-}}K^{-'}_{1}(0)
+ {\tr_{0} K^{+'}_{0}(0)\ H_{N 0} - i \tr_{0} K^{+}_{0}(0)\ H_{N 
0}^{2}\over \tr K^{+'}(0)} \,,
\label{secondderiv}
\ee
where $H_{n\,, n+1}={\cal P}_{n\,, n+1} R'_{n\,, n+1}(0)$, and we have
made use of the facts $K^{-}(0)=\sinh \xi_{-} \id$ and 
$\tr_{0} K^{+}_{0}(0)\ H_{N 0} = \tr K^{+''}(0)=0$.  
By explicitly evaluating (\ref{secondderiv}), we
obtain the Hamiltonian (\ref{Hamiltonian}).  Notice that the
Hamiltonian is Hermitian if $\xi_{\pm}$ are imaginary and $\kk_{\pm}$
are real. The corresponding energy eigenvalues $E$ are given by
\be
E = {\Lambda''(0)\over 8 (-1)^{N+1} \sinh \xi_{-}\sinh \xi_{+}} \,,
\label{energy}
\ee
where $\Lambda(u)$ are eigenvalues of the transfer matrix.

The transfer matrix has the periodicity property
\be
t(u + i \pi)= t(u) \,,
\label{periodicity}
\ee
as follows from (\ref{periodicityR}), (\ref{periodicityK}). Moreover,
the transfer matrix has crossing symmetry
\be
t(-u - {i \pi\over 2})= t(u) \,,
\label{transfercrossing}
\ee
which can be proved using a generalization of the methods developed in 
the appendices of \cite{MN2}. Finally, we note that the transfer 
matrix has the asymptotic behavior (for $\kk_{\pm} \ne 0$)
\be
t(u) \sim \kk_{-}\kk_{+} i^{N} {e^{u(4 + 2N)}\over 2^{1+2N}} \id + 
\ldots \qquad \mbox{for} \qquad
u\rightarrow \infty \,.
\label{transfasympt}
\ee

\section{Inversion identity}\label{sec:inversion}

Our main objective is to determine the eigenvalues $\Lambda(u)$ of the 
open-chain transfer matrix (\ref{transfer}), from which the energy 
eigenvalues (\ref{energy}) immediately follow.  We shall accomplish 
this using an exact inversion identity, which we first derive.
A similar approach was used in \cite{Ba, Fe} for closed chains.
This approach is based on the concept of fusion \cite{fusion, KS}.

The derivation of the inversion identity for the open XX spin chain/ 
6-vertex free-Fermion model closely parallels the one for the 8-vertex 
free-Fermion model considered in \cite{AN}.  For brevity, we shall 
often refer to these two models as the ${\cal N}=2$ and ${\cal N}=1$ 
models, respectively.  The principal tool which we use to derive the 
inversion identity is the open-chain fusion formula obtained in \cite{MN1}.  
We shall henceforth refer to this reference as I.

The matrix $R_{12}(u)$ at $u=-\rho={i \pi\over 2}$ is proportional to the 
one-dimensional projector $P_{12}^{-}$
\be
P_{12}^{-} = {1\over 2}  \left( \begin{array}{rrrr}
  0  &0  &0  &0  \\ 
  0  &1  &1  &0  \\
  0  &1  &1  &0 \\
  0  &0  &0  &0
\end{array} \right) \,, \qquad 
(P_{12}^{-})^{2} = P_{12}^{-} \,.
\ee 
As explained in I, from the corresponding degeneration of the
(boundary) Yang-Baxter equation, one can derive identities which allow
one to prove that {\it fused} (boundary) matrices satisfy {\it
generalized} (boundary) Yang-Baxter equations.

The fused $R$ matrix is given by (I 2.13)
\be
R_{<12> 3}(u) = P_{12}^+\ R_{13}(u)\ 
R_{23}(u + \rho)\ P_{12}^+ \,,
\ee 
where $P_{12}^{+} = \id  - P_{12}^{-}$. An important 
observation is that the fused $R$ matrix can be brought to the 
following upper triangular form by a similarity transformation
\be
X_{12}\ R_{<12> 3}(u)\ X_{12}^{-1} = \left( \begin{array}{cccc}
  s \id  &*             &*               &* \\ 
  0      &t \sigma^{z}  &-2t \sigma^{z}  &0  \\
  0      &0             &-t  \sigma^{z}  &0 \\
  0      &0             &0               &0
\end{array} \right) \,,
\ee
where 
\be
s=\cosh^{2}u \,, \qquad t=i \cosh u \sinh u \,,
\ee
and the $4 \times 4$ matrix $X$ is given by
\be
X = \left( \begin{array}{cccc}
  0  &1  &0  &0 \\ 
  1  &0  &0  &1  \\
  0  &0  &0  &1 \\
  0  &1  &1  &0
\end{array} \right) \,.
\ee 
It follows that the fused monodromy matrices (I 4.7), (I 5.4), (I 5.5)
\be
T_{<12>}(u) &=& R_{<12> N}(u) \cdots R_{<12> 1}(u) \,, 
\non \\
\hat T_{<12>}(u + \rho) &=& R_{<12> 1}(u) \cdots 
R_{<12> N}(u) \,,
\ee
also become triangular by the same transformation,
\be
X_{12}\ T_{<12>}(u)\ X_{12}^{-1} &=& 
\left( \begin{array}{cccc}
  s^{N} \id  &*             &*               &* \\ 
  0          &t^{N} F       &((-1)^{N}-1) t^{N} F  &0  \\
  0          &0             &(-t)^{N}  F     &0 \\
  0          &0             &0               &0
\end{array} \right) \non \\
&=& X_{12}\ T_{<12>}(u + \rho) \ X_{12}^{-1} \,,
\label{Tfusedtrasf}
\ee
where $F= \prod_{i=1}^{N}\sigma_{i}^{z}$.

The corresponding fused $K$ matrices are given by (I 3.5), (I 3.9)
\be
K^-_{<12>}(u) &=& P_{12}^+\ K^-_1(u)\ 
R_{12} (2u + \rho)\ K^-_2(u + \rho)\ P_{12}^+ 
\,, \non \\
K^+_{<12>}(u) &=&
\{ P_{12}^+\ K^+_1(u)^{t_1}\  R_{12} (-2u -3\rho)\ 
K^+_2(u + \rho)^{t_2} P_{12}^+ \}^{t_{12}} \,,
\ee 
since $M = V^{t} V = \id$.

Unlike the ${\cal N}=1$ case \cite{AN}, the similarity transformation 
does {\it not} bring also the fused $K$ matrices to upper triangular 
form.  \footnote{It is not possible to simultaneously triangularize 
both $R_{<12> 3}(u)$ and $K^-_{<12>}(u)$, since their commutator is 
not nilpotent; i.e., $\left[ R_{<12> 3}(u) \,, K^{-}_{<12>}(u) 
\right]^{n} \ne 0$.  A monograph on the general problem of 
simultaneous triangularization has recently been published 
\cite{RR}.} Nevertheless, the transformed fused $K$ matrices are 
``almost'' triangular
\be
X_{12}\ K_{<12>}^{\mp}(u)\ X_{12}^{-1} =  
\left( \begin{array}{cccc}
  m_{1}^{\mp}  &*        &*      &* \\ 
  0      &m_{2}^{\mp}    &m_{5}^{\mp}  &0  \\
  0      &m_{4}^{\mp}    &m_{3}^{\mp}  &0 \\
  0      &0        &0      &0
\end{array} \right) \,,
\label{Kfusedtrasf1}
\ee
where
\be
m_{1}^{\mp}&=& \pm {1\over 2} \sinh 2u\left[ \cosh 2u -\cosh 2 \xi_{\mp}+
2 \kk_{\mp}^{2} \sinh^{2} 2u \right] \,, \non \\
m_{2}^{\mp}&=& - {i\over 2} \sinh 2u\left[ \sinh 2(u \pm \xi_{\mp}) -
8 \kk_{\mp}^{2} \cosh u \sinh^{3} u \right] \,, \non \\
m_{3}^{\mp}&=& - {i\over 2} \sinh 2u\left[ \sinh 2(u \mp \xi_{\mp}) +
8 \kk_{\mp}^{2} \cosh^{3} u \sinh u \right] \,, \non \\
m_{4}^{\mp}&=& i \kk_{\mp}^{2} \cosh 2u \sinh^{2} 2u  \,, \non \\
m_{5}^{\mp}&=& \pm {i\over 2} \sinh 4u \sinh 2\xi_{\mp} \,.
\label{Kfusedtrasf2}
\ee 

The fused transfer matrix $\tilde t(u)$ is given by (I 4.5), (I 4.6)
\be
\tilde t(u) = \tr_{12}\ K^+_{<12>}(u)\ 
T_{<12>}(u)\ K^-_{<12>}(u)\ 
\hat T_{<12>}(u + \rho) \,.
\ee 
With the help of the results (\ref{Tfusedtrasf}),
(\ref{Kfusedtrasf1}),(\ref{Kfusedtrasf2}), we now
obtain the remarkable result that the fused transfer matrix is
proportional to the identity matrix,
\be
\tilde t(u) = \tilde \Lambda(u) \id \,,
\label{remarkable}
\ee
where 
\be
\tilde \Lambda(u) &=& s^{2N} m_{1}^{+}m_{1}^{-} + t^{2N}\Big[
m_{2}^{+} m_{2}^{-} + m_{3}^{+} m_{3}^{-} 
+ (-1)^{N}(m_{4}^{+} m_{5}^{-} + m_{4}^{-} m_{5}^{+}) \non  \\
&+& (1-(-1)^{N})(m_{3}^{+} m_{4}^{-} + m_{3}^{-} m_{4}^{+} 
- m_{2}^{+} m_{4}^{-}- m_{2}^{-} m_{4}^{+} + 2 m_{4}^{-} m_{4}^{+})
\Big] \,.
\ee
A similar result also holds for the ${\cal N}=1$ case \cite{AN}.

The fusion formula is given by (I 4.17), (I 5.1)
\be
t(u)\ t(u+\rho) = {1\over \zeta( 2u + 2\rho)} 
\left[ \tilde t(u) + 
\Delta \left\{ K^+(u) \right\} \Delta \left\{ K^-(u) \right\}
\delta \left\{ T(u) \right\} \delta \left\{ \hat T(u) \right\} 
\right] \,, 
\label{fusion}
\ee 
where the transfer matrix $t(u)$ is given 
by (\ref{transfer}) (see also (I 4.1), (I 4.2)), and the 
quantum determinants \cite{IK, KS} are given by
(I 4.15), (I 5.3), (I 5.7)
\be
\delta \left\{ T(u) \right \} &=& \delta \left\{\hat T(u) \right\} 
= \zeta(u + \rho)^N \,, \non \\
\Delta \left\{ K^-(u) \right\} &=& 
\tr_{12} \left\{  P_{12}^- \ K^-_1(u)\ 
R_{12}(2u + \rho)\ K^-_2(u+\rho)\ V_1\ V_2\ \right\} \,, \non \\
\Delta \left\{ K^+(u) \right\} &=& 
\tr_{12} \left\{ P_{12}^- \ V_1\ V_2\ K^+_2 (u + \rho)\ 
R_{12}(-2u-3\rho)\  K^+_1(u) \right\} \,.
\label{qdeterminants}
\ee

It follows from (\ref{remarkable})-(\ref{qdeterminants}) that the transfer 
matrix obeys an exact inversion identity
\be
t(u)\ t(u-{i \pi\over 2}) = f(u) \id \,.
\label{inversionidentity1}
\ee
where the function $f(u)$ is given by
\be
f(u)=\tanh^{2} 2u \left[ g_{1}(u) \cosh^{4N}u + g_{2}(u) 
\sinh^{4N}u + g_{3}(u) \sinh^{2N}u \cosh^{2N}u \right] \,, 
\label{functionf}
\ee
with
\be
g_{1}(u) &=& {1\over 4}(\cosh 2u - \cosh 2\xi_{-} + 
2\kk_{-}^{2}\sinh^{2}2u)(\cosh 2u - \cosh 2\xi_{+} + 
2\kk_{+}^{2}\sinh^{2}2u) \non \\ 
&=& 16 \kk_{+}^{2} \kk_{-}^{2} \cosh(u + i \eta_{-})\cosh(u - i \eta_{-})
\cosh(u + i \eta_{+})\cosh(u - i \eta_{+})\non \\ 
& & \times \cosh(u + \vartheta_{-})\cosh(u - \vartheta_{-})
\cosh(u + \vartheta_{+})\cosh(u - \vartheta_{+}) \,, \non \\
g_{2}(u) &=& g_{1}(u + {i \pi\over 2}) \,, \non \\
g_{3}(u) &=& 2 \kk_{+}^{2} \kk_{-}^{2}\left[1+(-1)^{N}+\sinh^{2}2u 
\right]\sinh^{2}2u + (-1)^{N}\Big[ (\kk_{+}^{2}\cosh 2\xi_{-} 
+\kk_{-}^{2}\cosh 2\xi_{+} )\sinh^{2}2u \non \\
& &+ {1\over 2}(\cosh 2\xi_{-}\ \cosh 2\xi_{+}\ \sinh^{2}2u -
\sinh 2\xi_{-}\ \sinh 2\xi_{+}\ \cosh^{2}2u ) \Big] \non \\
&=& 2 \kk_{+}^{2} \kk_{-}^{2} \Big[ \sinh^{2}2u \cosh^{2} 2u
+(-1)^{N} \Big( \sin 2\eta_{-} \sinh 2\vartheta_{-}
\sin 2\eta_{+} \sinh 2\vartheta_{+} \cosh^{2} 2u \non\\
& & + \cos 2\eta_{-} \cosh 2\vartheta_{-}
\cos 2\eta_{+} \cosh 2\vartheta_{+} \sinh^{2} 2u \Big) \Big] \,.
\label{gees}
\ee

The inversion identity (\ref{inversionidentity1})-(\ref{gees}) is the 
first main result of our paper.  We have checked it numerically up 
to $N=3$.

\section{Eigenvalues and Bethe Ansatz Equations}\label{sec:eigenvalues}

Having obtained the inversion identity, we now use it to determine the 
eigenvalues of the transfer matrix.  The commutativity relation 
(\ref{commutativity}) implies that the transfer matrix has eigenstates 
$| \Lambda \rangle$ which are independent of $u$,
\be
t(u) | \Lambda \rangle = \Lambda(u) | \Lambda \rangle \,,
\label{eigenvalueproblem}
\ee
where $\Lambda(u)$ are 
the corresponding eigenvalues.  Acting on $|  \Lambda \rangle$ 
with the inversion identity, we obtain the 
corresponding identity for the eigenvalues
\be
\Lambda(u)\ \Lambda(u-{i \pi\over 2}) = f(u) \,.
\label{inversionidentity2}
\ee
Similarly, it follows from (\ref{periodicity}) and
(\ref{transfercrossing}) that the eigenvalues have the periodicity and
crossing properties
\be
\Lambda(u + i \pi) = \Lambda(u) \,, \qquad 
\Lambda(-u - {i \pi\over 2}) = \Lambda(u) \,.
\label{eigenvalueprops}
\ee
Finally, (\ref{transfasympt}) implies the asymptotic behavior
\be
\Lambda(u) \sim \kk_{-}\kk_{+} i^{N} {e^{u(4 + 2N)}\over 2^{1+2N}} + 
\ldots \qquad \mbox{for} \qquad
u\rightarrow \infty \,.
\label{asympt}
\ee

We shall assume that the eigenvalues have the form
\be
\Lambda(u) = \rho \sinh 2u \prod_{j=0}^{N}\sinh(u-u_{j}) \cosh(u+u_{j}) 
\label{Ansatz} \,,
\ee
where $u_{j}$ and $\rho$ are ($u$-independent) parameters which are to 
be determined.  Indeed, this expression satisfies the periodicity and 
crossing properties (\ref{eigenvalueprops}), and it has the correct 
asymptotic behavior (\ref{asympt}) provided that we set
\be
\rho = i^{N}4 \kk_{-}\kk_{+} \,.
\label{rho}
\ee 

We now substitute the Ansatz (\ref{Ansatz}) into the inversion identity
(\ref{inversionidentity2}), and obtain
\be
(-1)^{N} \rho^{2} \sinh^{2}2u \prod_{j=0}^{N}
{1\over 4}\sinh 2(u-u_{j}) \sinh 2(u+u_{j}) =f(u)
\,.
\ee
Recalling the explicit expression (\ref{functionf}),(\ref{gees}) for
$f(u)$, we verify that both sides of the equation have the same
asymptotic behavior $ \sim e^{u(8 + 4N)}$ for ${u\rightarrow \infty}$.
Since the LHS has zeros $\pm u_{j}$, these must be zeros of $f(u)$.
That is, \footnote{The three functions $g_{i}(u) \,, i= 1\,, 2\,, 3\,,$
are even functions of $u$. Hence, if $u_{j}$ is a root of $f(u)$, then 
so is $-u_{j}$.}
\be
g_{1}(u_{j}) \cosh^{4N}u_{j} + g_{2}(u_{j}) 
\sinh^{4N}u_{j} + g_{3}(u_{j}) \sinh^{2N}u_{j} \cosh^{2N}u_{j} = 0
\,.
\ee
Dividing by $\cosh^{4N}u_{j}$, we obtain
\be
g_{2}(u_{j}) \tanh^{4N}u_{j} + g_{3}(u_{j}) \tanh^{2N}u_{j} +
g_{1}(u_{j}) = 0 \,.
\label{rearranged}
\ee
Regarding (\ref{rearranged}) as the quadratic equation 
\be
g_{2} x^{2} + g_{3} x + g_{1} = 0 
\ee
in the variable $x = \tanh^{2N}u_{j}$, we conclude that the parameters
$u_{j}$ satisfy the Bethe Ansatz equations
\be
\tanh^{2N}u_{j} = {h(u_{j})\over g_{2}(u_{j})} \,,
\label{BAE}
\ee
where the function $h(u)$ is defined by
\be
h(u) = {-g_{3}(u) \pm \sqrt{g_{3}(u)^{2} - 4 g_{1}(u)\ g_{2}(u)}\over 2} 
\,.
\label{functionh}
\ee 

The square root in (\ref{functionh}) can be eliminated by making an
appropriate change of variables.  Indeed, with the help of
(\ref{gees}), one can show that
\be
h(u) &=& \kk_{+}^{2} \kk_{-}^{2} \Big\{ -\sinh^{2}2u \cosh^{2}2u
-(-1)^{N}\left( \gamma_{1} \cosh^{2} 2u + 
\gamma_{2} \sinh^{2} 2u\right) \non \\
& & \pm
\sinh 2u \cosh 2u \sqrt{\alpha \sinh^{2}2u + \beta} \Big\} \,,
\label{tough}
\ee
where
\be 
\alpha &=& {1\over 2}\left(\cos 4\eta_{-}+\cos 4\eta_{+}+\cosh 4\vartheta_{-}+
\cosh 4\vartheta_{+} \right) + (-1)^{N}\big[ 
\cos 2(\eta_{-}+\eta_{+}) \cosh 2(\vartheta_{-}-\vartheta_{+}) \non \\
& & +
\cos 2(\eta_{-}-\eta_{+}) \cosh 2(\vartheta_{-}+\vartheta_{+}) \big]
\,, \non \\
\beta &=& \left[ \sin 2\eta_{-} \sinh 2 \vartheta_{-} 
+ (-1)^{N} \sin 2\eta_{+} \sinh 2 \vartheta_{+} \right]^{2}
-{1\over 4}\big[ 
\sin 2(\eta_{-}+\eta_{+}) \sinh 2(\vartheta_{-}-\vartheta_{+})\non \\
& & +
\sin 2(\eta_{-}-\eta_{+}) \sinh 2(\vartheta_{-}+\vartheta_{+}) 
\big]^{2} \,, \non \\ 
\gamma_{1} &=& \sin 2\eta_{-} \sinh 2\vartheta_{-}
\sin 2\eta_{+} \sinh 2\vartheta_{+} \,, \non \\ 
\gamma_{2} &=& \cos 2\eta_{-} \cosh 2\vartheta_{-}
\cos 2\eta_{+} \cosh 2\vartheta_{+} \,.
\label{abc}
\ee 
Let us change from the
spectral parameter $u$ to the new spectral parameter $v$ defined by
\be
\sinh 2u = i \sn 2v \,, \qquad \cosh 2u = \cn 2v \,, 
\label{newvariablea}
\ee
where the modulus $k$ of the Jacobian elliptic functions is given
by
\be
k^{2}= {\alpha\over \beta} \,.
\label{modulus}
\ee
With the help of the identities  (see, e.g.,
\cite{WW}) 
\be
\cn^{2}z + \sn^{2}z = 1 \,, \qquad 
\dn^{2}z + k^{2} \sn^{2} z = 1 \,,
\ee
one can see that the function $h(u)$ can be reexpressed as
\be
h(u)=\kk_{+}^{2} \kk_{-}^{2} \Big\{ \sn^{2}2v \cn^{2}2v 
- (-1)^{N}\left(\gamma_{1} \cn^{2} 2v - 
\gamma_{2} \sn^{2} 2v \right)
\pm i \sn 2v \cn 2v \dn 2v \sqrt{\beta} \Big\} \,.
\label{elliptich}
\ee
Hence, the Bethe Ansatz solution (\ref{Ansatz}), (\ref{rho}), 
(\ref{BAE}) can be reformulated as
\be
\Lambda(u)=(-{1\over 2})^{N-1}\kk_{+} \kk_{-} \sn 2v \prod_{j=0}^{N}
\left(\sn 2v -\sn 2v_{j} \right) \,,
\label{eigenelliptic}
\ee
where the parameters $v_{j}$ satisfy
\be
\left( {\cn 2v_{j} - 1\over \cn 2v_{j} + 1}
\right)^{N}={\sn^{2}2v_{j} \cn^{2}2v_{j} 
- (-1)^{N}\left(\gamma_{1} \cn^{2} 2v_{j} - 
\gamma_{2} \sn^{2} 2v_{j} \right)
\pm i \sn 2v_{j} \cn 2v_{j} \dn 2v_{j} \sqrt{\beta} \over
(\cn 2 v_{j} - \cn 2i\eta_{-})(\cn 2 v_{j} - \cn 2\vartheta_{-})
(\cn 2 v_{j} - \cn 2i\eta_{+})(\cn 2 v_{j} - \cn 2\vartheta_{+})}
\,. \non \\
\label{BAEelliptic}
\ee

This Bethe Ansatz solution is the second main result of our paper.  
This result passes several tests.  Indeed, for $N=0,1$, the 
eigenvalues agree with those obtained by direct diagonalization of the 
transfer matrix.  Moreover, as discussed in the following section, our 
solution is similar to the known one \cite{ABBBQ, Sk} for the case of 
diagonal boundary terms.

\section{Special cases}\label{sec:special}

For generic values of the boundary parameters, the Bethe Ansatz 
solution presented in the previous section is formulated in terms of 
Jacobian elliptic functions.  However, for modulus $k=0$ or $k=1$, 
these elliptic functions degenerate into ordinary trigonometric or 
hyperbolic functions.  Equivalently, the argument of the square root 
in (\ref{tough}) then becomes a perfect square, and so the square root 
effectively disappears.  We now briefly consider some of these special 
cases.

\subsection{Diagonal case}

In the limit $\kk_{\pm} \rightarrow 0$, the $K$ matrices
(\ref{Kminusmatrix}), (\ref{Kplusmatrix}) become diagonal, and
therefore so do the boundary terms in the Hamiltonian
(\ref{Hamiltonian}).  The transfer matrix $t(u)$ now commutes with the
operator $F=\prod_{i=1}^{N}\sigma^{z}_{i}$, and hence, both operators can be
simultaneously diagonalized.  Denoting the corresponding eigenstates
by $| \Lambda^{(\pm)} \rangle$, we have
\be
t(u) | \Lambda^{(\pm)} \rangle &=& \Lambda^{(\pm)}(u) 
| \Lambda^{(\pm)} \rangle \,, \non  \\ 
F | \Lambda^{(\pm)} \rangle &=& \pm 
| \Lambda^{(\pm)} \rangle \,.
\ee
The transfer matrix now has the asymptotic behavior
\be
t(u) \sim \rho  {e^{u(2 + 2N)}\over 2^{1+2N}} F + 
\ldots \qquad \mbox{for} \qquad
u\rightarrow \infty \,,
\ee
where 
\be
\rho = \left\{ \begin{array}{ll}
       i \cosh(\xi_{+} - \xi_{-}) & \mbox{for} \quad $N=$ \
       \mbox{even} \\
      i \sinh(\xi_{+} - \xi_{-}) & \mbox{for} \quad $N=$ \
       \mbox{odd}
       \end{array} \right. \,.
\ee 
The eigenvalues are given by
\be
\Lambda^{(\pm)}(u) = \pm \rho \sinh 2u \prod_{j=1}^{N}
\sinh(u-u_{j}) \cosh(u+u_{j}) \,,
\ee
which is similar to (\ref{Ansatz}) except with one less root. 

In terms of the boundary parameters $\eta_{\pm} \,, \vartheta_{\pm}$
(\ref{alternative}), the limit $\kk_{\pm} \rightarrow 0$ corresponds
to
\be
\eta_{\pm} = i \xi_{\pm} - {\pi\over 2} \,, 
\qquad e^{-\vartheta_{\pm}} = {1\over \kk_{\pm}} \rightarrow \infty \,.
\label{limit}
\ee 
In this limit the function $h(u)$ becomes equal (for $N=$ even) to
\be
h(u) = -\sinh(u \mp \xi_{-})\cosh(u \mp \xi_{-})
\sinh(u \pm \xi_{+})\cosh(u \pm \xi_{+}) \,.
\ee 
Moreover, the Bethe Ansatz equations (\ref{BAE}) become
\be
\tanh^{2N}u_{j} = 
-{\sinh(u_{j} \mp \xi_{-})\sinh(u_{j} \pm \xi_{+})\over 
\cosh(u_{j} \pm \xi_{-})\cosh(u_{j}\mp \xi_{+})} \,.
\ee
These results are similar to those obtained previously \cite{ABBBQ,
Sk} for the XXZ chain.  The two choices of signs correspond to the two
possible pseudovacua -- either all spins up or all spins down.

\subsection{Nondiagonal cases}

Within the space of boundary parameters, there are various 
submanifolds, such as
\be
\eta_{-}-\eta_{+} \pm i(\vartheta_{-}-\vartheta_{+}) = 
{\pi\over 2} {1 + (-1)^{N}\over 2} \,,
\ee
for which $\alpha=k=0$.

As a simple example, let us consider the particular case
$N=$ odd, $\eta_{-}=\eta_{+} \equiv \eta $,
$\vartheta_{-}=\vartheta_{+} \equiv \vartheta$, for which also 
$\beta =0$.  The function $h(u)$ becomes equal to
\be
h(u)=-\kk_{+}^{2}\kk_{-}^{2}\sinh 2(u - i\eta)\sinh 2(u + i\eta)
\sinh 2(u - \vartheta) \sinh 2(u + \vartheta) \,.
\ee
We then obtain the Bethe Ansatz equations 
\be
\tanh^{2N}u_{j} = 
-\coth(u_{j} + i \eta)\coth(u_{j} - i \eta)
\coth(u_{j} + \vartheta)\coth(u_{j} - \vartheta) 
\,.
\ee

\section{Discussion}\label{sec:discuss}

This work raises a number of interesting questions, some of which we 
list below:

We have seen that the doubly-periodic functions in the Bethe Ansatz 
solution degenerate into singly-periodic functions for special values 
of the boundary parameters.  In \cite{Bi}, important simplifications 
are also found to occur for special values of the boundary parameters.  
It is likely that these two observations are related.

As we have emphasized, an exact inversion identity for the XX (or
${\cal N}=2$) case is made possible by the key fact that the fused
transfer matrix is proportional to the identity matrix.  A similar
result also holds for the ${\cal N}=1$ case \cite{AN}.  It would be
interesting to better understand the relation of this phenomenon to
the free-Fermion condition.

The model (\ref{Hamiltonian}) is not the most general integrable open 
XX chain. Indeed, the most general solution \cite{dVGR, GZ} of the 
XXZ boundary Yang-Baxter equation has the off-diagonal terms 
$\kk^{(1)}_{\pm}\sinh 2u$ and $\kk^{(2)}_{\pm}\sinh 2u$, while here we have 
restricted to the special case 
$\kk^{(1)}_{\pm}=\kk^{(2)}_{\pm} \equiv \kk_{\pm}$.
(See Eqs. (\ref{Kminusmatrix}),  (\ref{Kplusmatrix}).) However, we do 
not expect that the more general case will lead to new significant 
complications. In particular, we expect that the same approach can be 
used to derive an exact inversion identity and to obtain the 
corresponding Bethe Ansatz solution.

Since the model (\ref{Hamiltonian}) has various boundary parameters,
its phase diagram is likely to have a rich structure.  Our exact Bethe
Ansatz solution should provide a means of exploring these phases. 
Moreover, as mentioned in the Introduction, this solution opens the
way to formulating the thermodynamic Bethe Ansatz equations for
integrable ${\cal N}=2$ supersymmetric quantum field theories with
boundary \cite{FI, Ne}, as was done for the case of ${\cal N}=1$
supersymmetry in \cite{AN}.  Finally, with the insight gained from the
XX chain, it might now be possible to finally solve the open XXZ chain
with nondiagonal boundary terms.

We hope to report on some of these problems in future publications.

\section*{Acknowledgments}

I thank O. Alvarez for helpful discussions.
This work was supported in part by the National Science Foundation
under Grants PHY-9870101 and PHY-0098088.

\end{document}